\documentclass[preprint,12pt,authoryear]{elsarticle}

\usepackage{amssymb}
\usepackage[normalem]{ulem}
\usepackage{color}
\usepackage{siunitx}

\usepackage{lineno}

\journal{XXXXXXX}

\biboptions{authoryear}

\begin{document}

\begin{frontmatter}



\title{Charge injection into the atmosphere by explosive volcanic eruptions through triboelectrification and fragmentation charging}


\author[1]{Joshua M\'endez Harper}
\author[2]{Corrado Cimarelli}
\author[2]{Valeria Cigala}
\author[2]{Ulrich Kueppers}
\author[1]{Josef Dufek}

\address[1]{Department of Earth Science, University of Oregon, Eugene OR, 97403, USA}
\address[2]{Department f\"ur Geo- und Umweltwissenschaften, Ludwig-Maximilians-Universit\"at M\"unchen, Munich BY, 80333, Germany}

\begin{abstract}

Volcanic eruptions are associated with a wide range of electrostatic effects. Increasing evidence suggests that high-altitude discharges (lightning) in maturing plumes are driven by electrification processes that require the formation of ice (analogous to processes underpinning meteorological thunderstorms). However, electrical discharges are also common at or near the volcanic vent. A number of ``ice-free" electrification mechanisms have been proposed to account for this activity: fractocharging, triboelectric charging, radioactive charging, and charging through induction. Yet, the degree to which each mechanism contributes to a jet's total electrification and how electrification in the gas-thrust region influences electrostatic processes aloft remains poorly constrained. Here, we use a shock-tube to simulate overpressured volcanic jets capable of producing spark discharges in the absence of ice. These discharges may be representative of the continual radio frequency (CRF) emissions observed at a number of eruptions. Using a suite of electrostatic sensors, we demonstrate the presence of size-dependent bipolar charging (SDBC) in a discharge-bearing flow for the first time. SDBC has been readily associated with triboelectric charging in other contexts and provides direct evidence that contact and frictional electrification play significant roles in electrostatic processes in the vent and near-vent regions of an eruption. Additionally, we find that particles leaving the region where discharges occur remain moderately electrified. This degree of electrification may be sufficient to drive near-vent lightning higher in the column. Thus, near-vent discharges may be underpinned by the same electrification mechanisms driving CRF, albeit involving greater degrees of charge separation. 

\end{abstract}


\end{frontmatter}

\section{Introduction}

\subsection{Plume lightning and proximal discharges}

Investigations over the last two decades reveal that electrical activity in volcanic columns may be broadly characterized into plume lightning and vent/near-vent discharges \citep{thomas_electrical_2007, behnke_observations_2013,  cimarelli_multiparametric_2016, aizawa_physical_2016, van2020did}. The first modality comprises large-scale discharges at elevation in maturing plumes and, in many regards, is analogous to meteorological lightning \citep{prata2020anak, van2020did}. Because of the large energies involved, plume lightning can often be detected with wide-range lightning networks (\cite{van2020did, prata2020anak}). The second category, vent and near-vent discharges, are electrical events that neutralize lower amounts of charge per event and, as their names suggest, occur closer to the volcanic vent \citep{thomas_electrical_2007, behnke_observations_2013, behnke2018investigating}. Although often lumped together, \cite{behnke2018investigating} showed that vent and near-vent discharges originate from fundamentally distinct breakdown processes. Vent discharges are innumerable streamer discharges that occur within or directly above the vent and are no more than a few tens of meters in length \citep{thomas_electrical_2007, behnke_observations_2013, behnke2018investigating}. With current measurement techniques, these minute discharges cannot be detected individually (either at optical or RF wavelengths) or at very great distances. Collectively, however, vent discharges produce a continuous electromagnetic ``hum" (commonly referred to as continual radio frequency or CRF) that can be observed with instruments like the lightning mapping array \citep{thomas_electrical_2007, behnke_observations_2013, behnke_changes_2015, behnke2018investigating}. CRF is often detected together with seismic and acoustic signals implying a relationship with explosions and over-pressure conditions at the vent (Note: we will use vent-discharges and CRF sources interchangeably) \citep{smith2020examining}. Occurring somewhat higher in the column and  later in the eruption, near-vent lightning involves leader discharges that can have lengths between a few hundred meters to several kilometers and can be detected individually \citep{aizawa_physical_2016,cimarelli_multiparametric_2016, behnke2018investigating}. Although larger than vent discharges, near-vent lightning still moves smaller amounts of charge per event than meteorological lightning and, thus, may be invisible to global detection networks \citep{vossen2021long}. Locally, however, it may produce changes to the ambient electric field \citep{behnke2018investigating}. \cite{aizawa_physical_2016} notes that meteorological/plume lightning shares many characteristics with near-vent lightning, hinting that separating both into two categories may be unnecessary. Nonetheless, an explicit distinction between the two (which we make in the present work) may be warranted given the likely differences in electrification mechanisms underlying near-vent and plume/meteorological lightning.  

\begin{figure}[h]
	\centering
	\includegraphics[width=\textwidth]{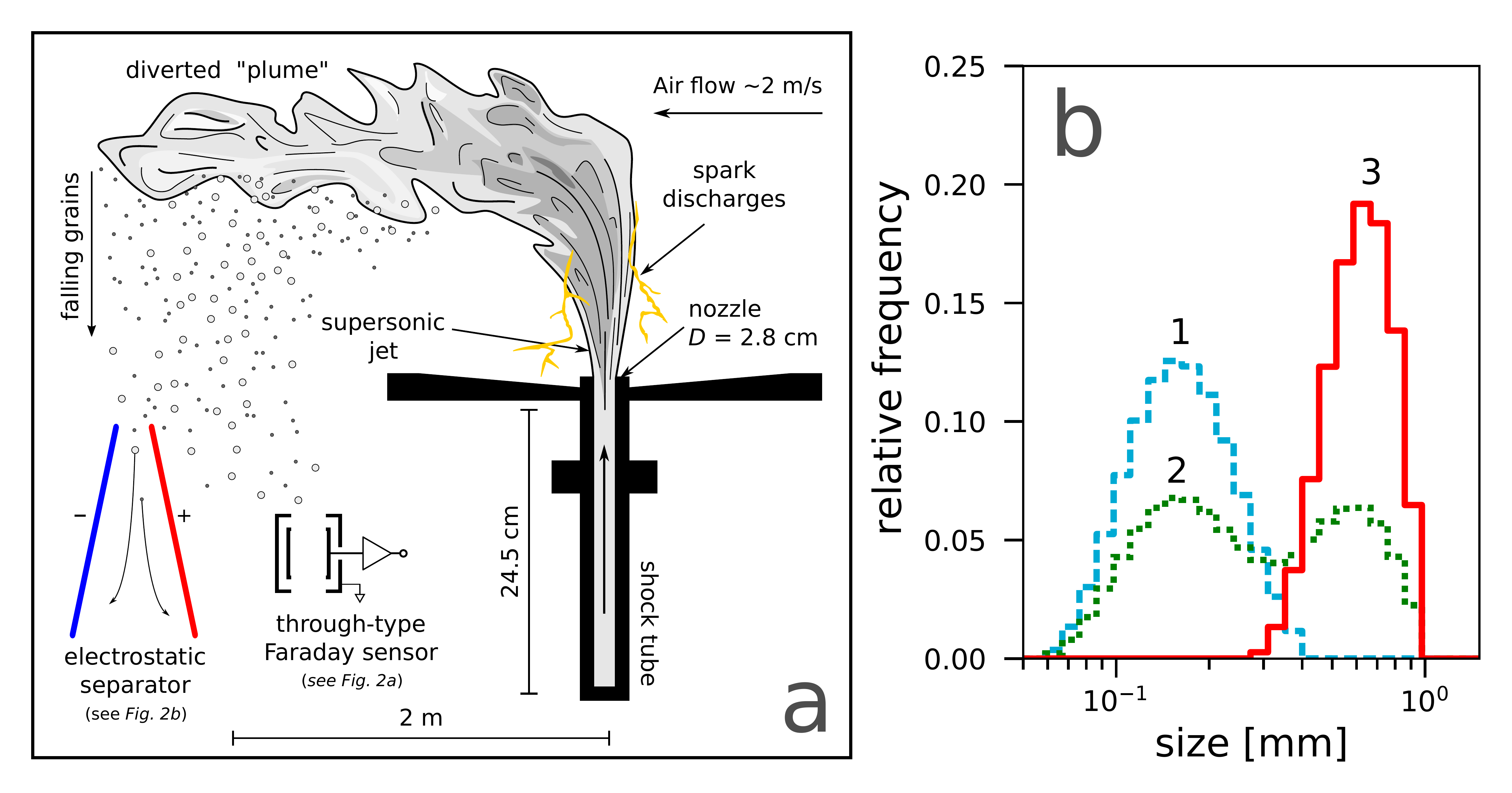}
	\caption{Experimental setup. a) We generate granular jets with the shock-tube setup described previously in \cite{cimarelli_experimental_2014}. The jet's dynamics approximate those in the gas-thrust region of volcanic columns and produce spark discharges. Venting into free air, the ``column" is then diverted away from the shock-tube by an air stream. Particles falling out of the ``plume" are then characterized by two electrostatic measurement systems: 1) an electrostatic separator and 2) a through-type Faraday sensor (described in \textbf{Figure \ref{setup2}a} and \textbf{b}, respectively). b) Size distributions for the three samples used in these experiments as measured by a diffraction analyzer with the following nominal ranges: 1) 90-300 $\mu$m, 2) 50$\%$ 90-300 and 50$\%$ 300-1000 mixture $\mu$m (by volume), and 3) 300-1000 $\mu$m. }
	\label{setup1}
\end{figure}

\begin{figure}
	\centering
	\includegraphics[width=0.6\textwidth]{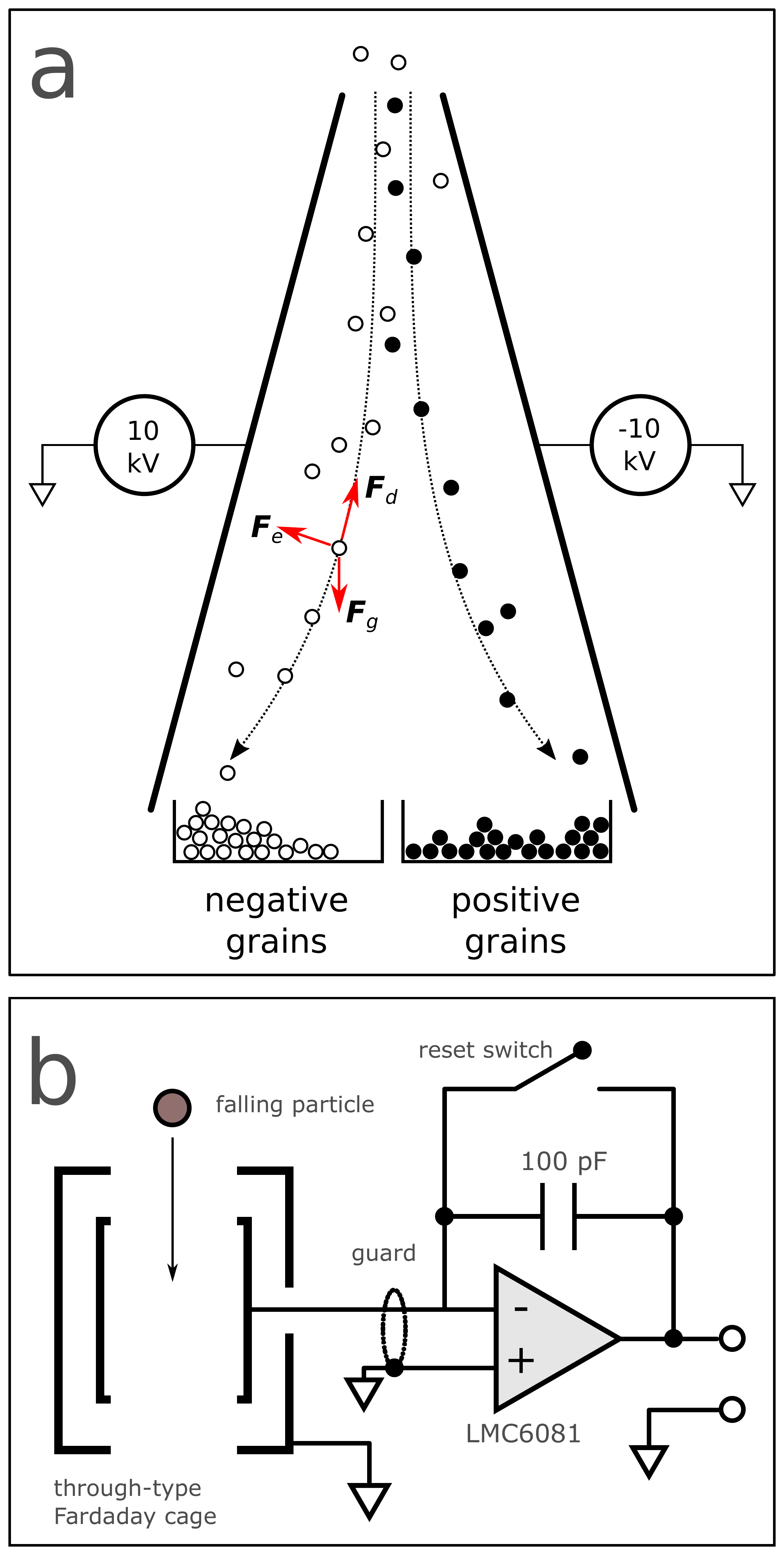}
	\caption{
	Detail of charge characterization subsystems referred to in \textbf{Figure \ref{setup1}}. a) Electrostatic separator to asses the charge polarity on particles as a function of size. Settling grains fall between two plates with a potential difference of 20 kV between them. The electric field separates negative and positive particles and these are collected in two bins at the bottom of the separator. Grains in the bins are then sized using an optical microscope and ImageJ. In this manner, we can investigate how negative and positive charge is partitioned on particles based on size. b) Charge magnitude measurement subsystem. The charge on particles falling out of the ``plume" are measured by an array of 8 miniature through-type Faraday cups (TTFC). Each sensor is capable of resolving charges as small as $<$10 femto-Coulombs (fC).  See \citep{mendez_harper_effects_2016, Mendez2018KCL} for more details.} 
	\label{setup2}
\end{figure}

An ever growing number of observations suggests that vent and near-vent discharges --what we will collectively call proximal discharges-- are common during explosive eruptions \citep{thomas_electrical_2007, behnke_observations_2013, aizawa_physical_2016, cimarelli_multiparametric_2016, behnke2018investigating, smith2020examining, vossen2021long}. These observations imply that erupted material charges efficiently within the conduit and in the jet-thrust region. Furthermore, there is evidence that proximal discharges contain valuable information about the source of the eruption. For instance, CRF is only detected with forcing at the vent and occurs within the gas-thrust region \citep{behnke2018investigating, smith2020examining}. \cite{smith2020examining} demonstrated this fact by showing that CRF can be correlated with the acoustic and seismic signals associated with active fragmentation. Experimentally, \cite{Mendez2018Infer} showed that the location and timing CRF emissions reflect the geometry and temporal evolution of barrel shock structures in supersonic jets.  These spatiotemporal constraints suggest CRF is a valuable tool to detect incipient eruptions \citep{behnke2018investigating}.  Regarding near-vent lightning, \cite{cimarelli_multiparametric_2016} indicate that the number of discharges is proportional to the over-pressure at the vent. These authors conclude that the intensity of near-vent electrical activity scales with the energy of eruptions. Furthermore, \cite{aizawa_physical_2016} argue that the volumetric charge density in proximal jets may be much larger than that in thunderstorms. Because of these elevated charged loadings, the proximal regions of the volcanic system may also be interrogated using active methods such as GNSS occultation \citep{mendez2019effect}. Using an array of electrostatic instruments, \cite{behnke2018investigating} report complex feedback mechanisms between CRF sources and larger near-vent discharges, suggesting that both forms of discharge may depend on a shared charge budget.

Nonetheless, the physical, chemical, and dynamical processes that charge pyroclasts within the conduit and the gas-thrust region remain poorly constrained. Ice and graupel are generally  absent in any large quantities \citep{cimarelli_multiparametric_2016, vossen2021long}. Thus, in contrast to volcanic lightning at altitude \citep{van2020did, prata2020anak}, electrification mechanisms comparable to those in thunderclouds cannot account for electrical activity near the vent. Instead, proximal discharges likely reflect ``dry" charging processes operating with varying degrees of efficiency within the conduit and an expanding jet.

Starting at depth, material possibly charges during the brittle failure of the magmatic column and subsequent disruptive clast-clast collisions \citep{james2000volcanic}. Fractocharging may involve a number of pathways, including piezoelectricity, pyroelectricity, atomic dislocations, positive-hole activation, and the release and capture of positive and negative ions as new surfaces are created \citep{dickinson1981emission, xie2018triboluminescence}. \cite{james2000volcanic} fractured pumice through repeated impacts and abrasion and found that fragments carried elevated surface charge densities. Quite recently, \cite{smith2018correlating} found that eruptions producing more equant grains were associated with CRF, perhaps suggesting that milling (secondary fragmentation) plays a role in vent-discharges. It is worth noting that, although the fracture mechanism is often invoked to account for electrification in the near-vent region, not a single experimental follow up work has been conducted on the matter using natural materials (pumice) in the last 20 years. As such, fractocharging is perhaps the least-well understood ``major" charging mechanism in the volcanic context.

Non-disruptive collisions may too lead to electrification through the well-known (but imperfectly understood) triboelectric effect \citep{hatakeyama_disturbance_1951, kikuchi_atmospheric_1982, aplin_electrical_2014, mendez_harper_effects_2016, mendez2017electrification, mendez2020microphysical, mendez2021detection}. Importantly, not only does triboelectricity have the ability to produce efficient charging in a granular material, but may separate charges of opposite polarity based on particle size \citep{hatakeyama_disturbance_1951, kikuchi_atmospheric_1982, zhao2003bipolar, forward_particle-size_2009, waitukaitis2014size}. Indeed, triboelectric charging often results in smaller, negatively-charged grains and larger grains with generally positive charges. This phenomenological feature,  \textit{size-dependent bipolar charging} (SDBC), may be critical in the production of discharges in proximal volcanic jets (and other dusty planetary environments) as particles of different sizes and opposite charge become separated through hydrodynamics \citep{cimarelli_experimental_2014} or sedimentation \citep{harrison2016applications}. 

A handful of studies have been explicitly designed to investigate triboelectric SDBC using volcanic materials. \cite{forward_particle-size_2009} employed a fluidized bed to electrify basalt particles under partial vacuum. This study, however, used heavily altered materials to approximate Martian regoliths rather than recently erupted ash. Nonetheless, serendipitous reports of size-dependent bipolar charging in chemically-unmodified volcanic ash exist in the literature. Many of these observations were not placed within the modern framework of triboelectrification simply because the models had not yet been formulated. \cite{hatakeyama_disturbance_1951} studied the frictional electrification of Aso and Asama ash samples. Those investigators reported  standard SDBC --that is positive large grains, negative small grains-- in Aso ash. However, the Asama ash samples displayed inverse SDBC (negative large grains, positive small grains). In these experiments, particles were allowed to contact foreign objects (an aluminum plate, for example), possibly biasing the polarity of the charge in manners that would not be encountered in natural systems. Thirty years later, \cite{kikuchi_atmospheric_1982} conducted similar experiments and found standard SDBC in ash particles from the 1977 Usu eruption. At Sakurajima, \cite{miura2002measurements} measured changes in the atmospheric potential gradient associated with small explosive events and estimated the surface charge density and polarity of ash falling out of plumes using an electrostatic separator (a method similar to the one we describe below). Those authors report particles with surface charge densities approaching the ionization limit (10\textsuperscript{-6} - 10\textsuperscript{-5} Cm\textsuperscript{-2}) and  standard SDBC. 

In addition to tribo- and fractoelectric processes, other mechanisms have been proposed to account for proximal discharges. \cite{pahtz_why_2010} suggests that materials like volcanic ash and mineral dust could charge through the polarizing effects of an ambient electric field. \cite{aplin_electrical_2014} provide evidence that the decay of radioactive elements in the magma may lead to ``self-charging" of ash. Very recently, \cite{nicoll2019first} deployed sensors into a plume at Stromboli, finding that the gas phase itself is charged. 

Building monitoring tools that effectively leverage proximal electrical effects requires a better understanding of the mechanisms that charge pyroclasts. Accomplishing such a feat, however, is complicated by the fact that much uncertainty remains regarding proposed electrification mechanisms themselves. For instance, while triboelectrification has been described since the time of the ancient Greeks, we have yet to unequivocally identify the charge carriers being exchanged during frictional interactions \citep{lacks_contact_2011, lacks2019long}. These charge carriers could be electrons, ions, or both. Similarly, some authors have presented evidence that triboelectrification arises from surface damage at minute spatial scales, implying that contact and frictional electrification are ultimately forms of fragmentation charging \citep{pan2019fundamental, lacks2019long}.

\begin{figure}
	\centering
	\includegraphics[width=1.1\textwidth]{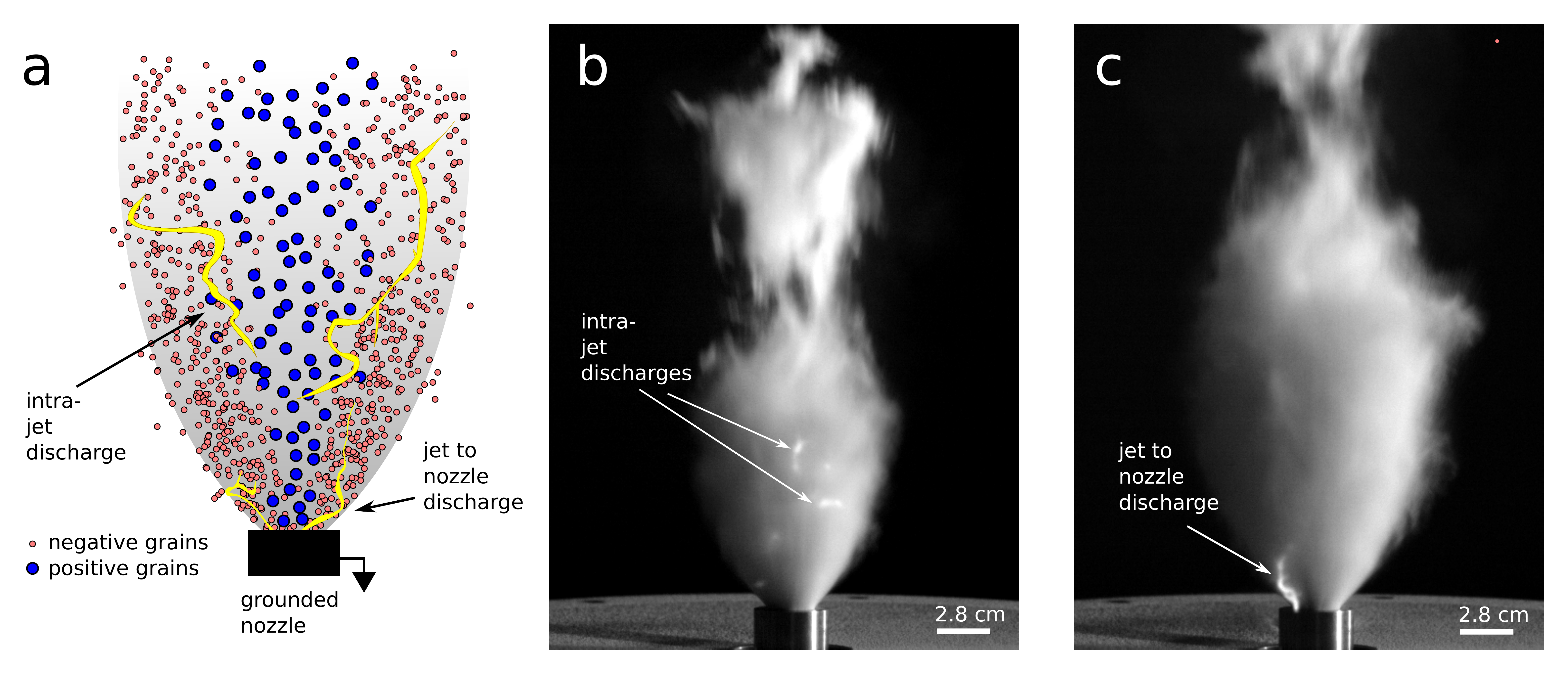}
	\caption{a) Schematic showing two kinds of discharge in the experiments. 1) Intra-jet discharges occur between grains or clusters of grains of opposite polarity. 2) Jet to nozzle discharges occur between the charged grains and grounded metallic nozzle. b) Typical intra-jet discharges in an experiment with the 90-300 $\mu$m particle size distribution. c) Jet to nozzle discharge in the same experiment as b).}
	\label{discharge}
\end{figure}

Beyond questions surrounding the charging mechanisms that putatively drive vent and near-vent discharges, little is known about how proximal electrification influences the long term electrostatic evolution of the eruptive column. One possibility is that pyroclasts advected high into the atmosphere retain charge generated in the conduit and the gas-thrust region. This ``pre-charging" may have important consequences for subsequent electrical effects, as some work indicates that charging in a granular material depends on pre-existing electric fields (e.g. \cite{pahtz_why_2010}). A second possibility is that the abundance of proximal discharges effectively neutralizes charge gained at or near the vent. Recombination in the gas-thrust region would imply that ``downstream" lightning storms in mature plumes generally necessitate additional cycles of electrification (perhaps driven by ice) and reflect little about eruption dynamics at the source. Evidence for this second hypothesis exists in field data collected at Augustine \citep{thomas_electrical_2007} and Redoubt \citep{behnke_observations_2013}, which show that electrical activity waned after the initial explosive phases. These periods of electrical inactivity could signify that volcanic columns emerge from the gas-thrust region with weak degrees of charging. The resumption of electrical activity in mature plumes could indicate activation of water-based electrification mechanisms \citep{prata2020anak, van2020did}.      

Here, we use a shock-tube to simulate explosive, overpressured volcanic jets and address a subset of the questions posed above. Our setup allows us to investigate the charge mechanisms that drive CRF sources and make inferences regarding subsequent near-vent lightning. For the first time, we identify nominal size-dependent bipolar charging in a simulated volcanic jet bearing streamer discharges. SDBC in our shock-tube experiments provides direct evidence that tribocharging is a dominant electrification mechanism in the gas-thrust region. Additionally, we find that particles emerging from the supersonic flow carry charge densities comparable to those measured on grains falling out of proximal volcanic columns (e.g. \cite{gilbert_charge_1991, miura2002measurements}). Further analysis shows that this amount of charge may be sufficient to drive near-vent lightning. As such, our results indicate that near-vent lightning is likely underpinned by the same electrification mechanisms as CRF sources, but reflects larger scale charge separation in columns. 

\section{Methodology}

We use the shock-tube setup described previously by \cite{cimarelli_experimental_2014} to produce artificial volcanic discharges (\textbf{Figure \ref{setup1}a}).  The shock-tube (24.5 cm in length, 2.8 cm in diameter) was loaded with approximately 75 ml of  volcanic ash and then pressurized to 10 MPa with argon gas. Exceeding this pressure ruptures a set of two diaphragms, causing the granular material to be ejected from the tube by explosive decompression. The dynamics of the decompression event and the resulting supersonic jet simulate those expected in the conduit and gas-thrust region of a volcanic jet \citep{cimarelli_experimental_2014, gaudin2019electrification}. This apparatus has the ability to charge pyroclasts through two principal mechanisms: fractoelectrification and tribocharging. The contributions from other putative electrification mechanisms such as radioactive decay or induction are excluded. 

Unlike previous efforts in which the shock-tube vented into a metallic collection chamber, we allowed the jet to expand into a large room to minimize collisions between grains and foreign surfaces. A low-powered air stream with a velocity of 2 m/s was generated perpendicular to the flow at a height of 1 m above the nozzle ($\sim$80 cm above the Mach disk). This artificial ``wind" was used to deflect lofted particles away from the ejection axis (see schematic in \textbf{Figure \ref{setup1}a}). Pushed away from the shock-tube and falling under the action of gravity, grains were sampled by two devices: 1) an electrostatic separator (ESS) and 2) an array of eight miniature through-type Faraday cages (TTFC) capable of measuring the charge on individual grains . Both the ESS and TTFC array were placed 1.5 m downwind of the nozzle. We note that a principal source of error associated with this methodology is that the size distribution of sampled particles may be different from that of the original sample. As discussed in \cite{cimarelli_experimental_2014}, experimental flows composed of larger particles are dominated by inertia. Thus, these particles were difficult to deflect using the artificial wind. Additionally, if successfully deflected, larger particles may fall out of the plume before they reach the sensors. We attempted to compensate for these effects by strategically placing the sensors in the proximity of the shock-tube, but we note that the constraints of the laboratory environment could not completely alleviate them. The overall result is that our measurement technique is better suited to characterize samples with fine grain size distributions. 

\begin{figure}
	\centering
	\includegraphics[width=0.6\textwidth]{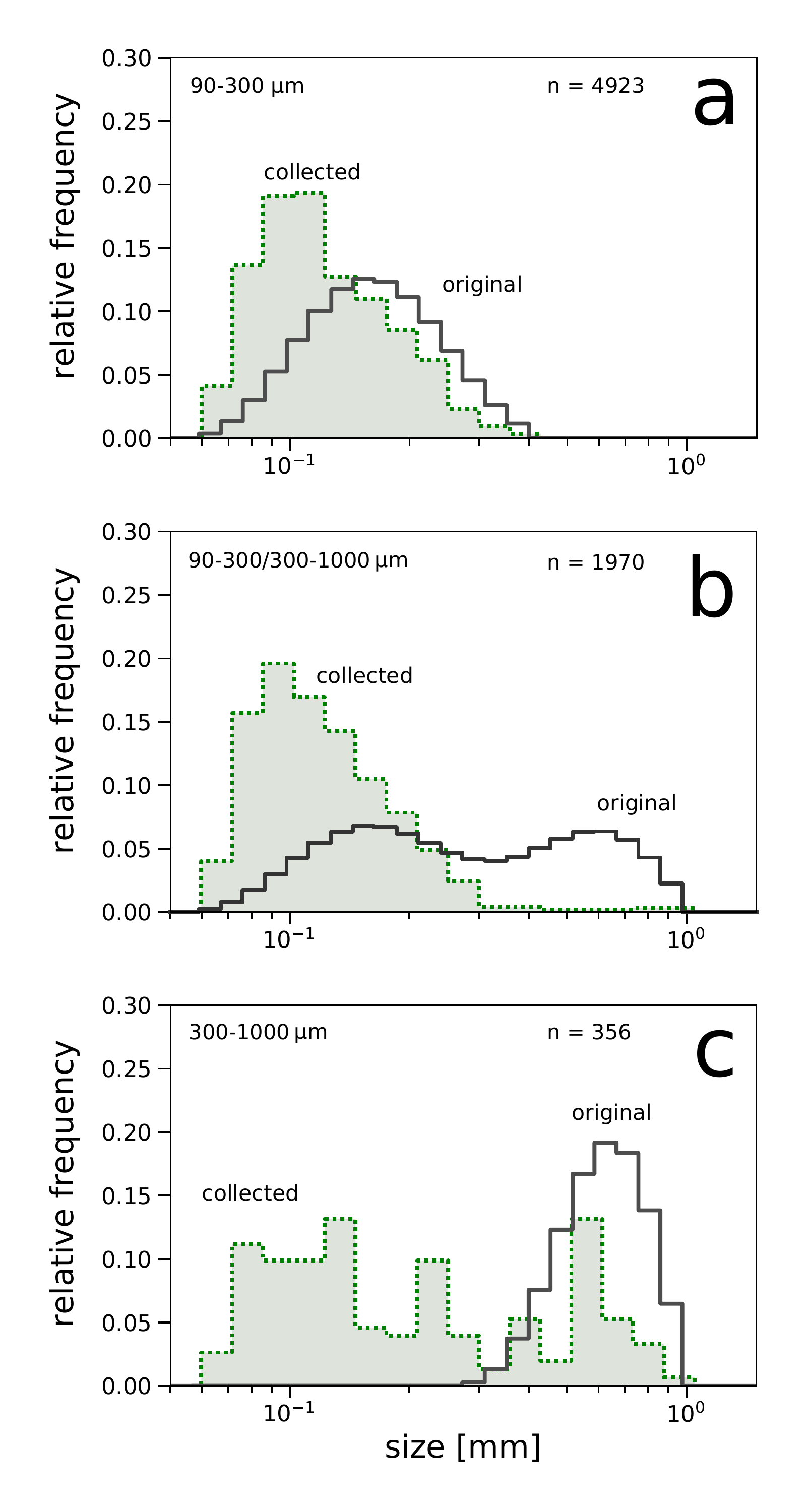}
	\caption{Particle size distributions for particles collected in the electrostatic separator (dotted green, filled) and original sample (solid black, unfilled); a) 90-300 $\mu$m; b) 90-300 and 300-1000 $\mu$m (at 1:1 weight ratio); and c) 300-1000 $\mu$m. The number of particles characterized per collected sample is indicated in the top right of each panel.}
	\label{histogramsO}
\end{figure}

The ESS consists of two vertical, sub-parallel, 1 m-long copper plates with a potential difference of 20 kV between them (See \textbf{Figure \ref{setup2}a}). When a charged particle passes between these plates, its trajectory is modified by the imposed electric field. Negatively-charged grains are diverted toward the positive plate, whereas grains carrying positive charge drift toward the negative plate. Thus, particles were separated by polarity and collected in two bins placed at the base of the separator. We then characterized the particles with an optical microscope to obtain spherical-equivalent diameters. 

The TTFC array was micromachined directly onto a printed circuit board. Each Faraday cup has an aperture of 1 mm and is capable of measuring charges  $<$10 fC. A simplified schematic for a single channel of the TTFC is shown in \textbf{Figure \ref{setup2}b}. When a particle traverses the sensing volume of one of the TTFC channels, an amplifying stage produces a voltage pulse whose magnitude is proportional to the charge on the particles. This output voltage was digitized by a National Instruments NI-6008 data acquisition unit and a PC running LabVIEW. 

Each shock-tube decompression event was recorded with a high-speed camera at a frame rate of 36,000 fps, allowing us to capture discrete spark discharges in the flow. We employed washed and sieved pumice quarried from deposits of the 13 ka Laacher See maar eruption in the Eifel Volcanic Field (Germany) with three different granulometries: fine (nominally 90-300 $\mu$m), coarse (nominally 300-1000 $\mu$m), and medium (a 50/50 mixture by volume of the fine and coarse samples). Particle size distributions before the experiments were obtained with a Coulter LS230 laser sizer. Because the number of particles sampled per experiment by both the ESS and the TTFC is relatively low, we repeated each experiment four times with each granulometry.

\section{Results and discussion}

\subsubsection{Electrical discharges}

All experimental jets produced electrical discharges.  A typical electrical spark in an experiment using the 90-300 $\mu$m sample is rendered in \textbf{Figure \ref{discharge}}. We observed discharges exclusively within the region of rarefying jet expansion as described by \cite{Mendez2018Infer} (we note that discharges could have also occurred within the shock-tube itself, but we were unable to image these). As reported by \cite{cimarelli_experimental_2014} and \cite{gaudin2019electrification}, we find that the number of discharges generally increased with the proportion of fines. Experiments with the fine granulometry displayed innumerable small discharges, whereas jets with coarse particles produced no more than a dozen discharges events per experiment. \cite{cimarelli_experimental_2014} suggest that these distinct behaviors underscore differences in gas-particle coupling between experiments. Specifically, in jets with abundant fines particles cluster in turbulent eddies. Conversely, in experiments with coarse grains, the motion of particles remains collimated by inertia. These dynamics depend on SDBC.

We clarify that the discharges we observe in our experiments are not representative of large scale lightning, volcanic or otherwise. The small sparks described here and in previous works (e.g. \cite{gaudin2019electrification, stern2019electrification}) are cold plasma channels or corona streamers, rather than highly conductive, hot leaders. Many of us have experienced the (startling) effects of streamers when reaching for a doorknob after scuffing our shoes on a carpeted floor. As noted above, \cite{behnke2018investigating} hypothesize that CRF may be the collective manifestation of many streamers occurring at the vent of an erupting volcano. We consider the discharges in our experiments to be laboratory analogs of these CRF-generating streamers.

\subsection{Grain size distributions}

Particles in the jet were diverted by a horizontal air stream toward an electrostatic separator and a Faraday cup array. We sized the grains that fell into the ESS using an optical microscope. The size distributions for each sample are rendered in \textbf{Figure \ref{histogramsO}} (green, filled curves). These histograms aggregate data from four shock-tube experiments. For reference, the pre-experiment size distributions as measured by a diffraction-based size analyzer are also displayed in \textbf{Figure \ref{histogramsO}} (unfilled, solid curves). 

By comparing the pre- and post-experiment histograms in \textbf{Figure \ref{histogramsO}}, we see that the experimental process modified the original size distributions in two principal ways. Firstly, as discussed above, smaller particles were diverted in greater number than larger ones by the stream of air. Thus, the material collected in the ESS was biased toward the smaller grains. Such effect was more noticeable for samples containing particles in the 300-1000 $\mu$m size range (\textbf{Figure \ref{histogramsO}c} and \textbf{b}). Furthermore, the total number of particles collected by the separator decreased as the abundance of large particles increased (we report the total number of particles sampled in the upper right corners of panels in \textbf{Figure \ref{histogramsO}c}). A second effect, evident in \textbf{Figure \ref{histogramsO}}, was the presence of particles smaller than the smallest grain in the original sample. This implies that some amount of material was disrupted during the decompression event and/or subsequent transport \citep{dufek_granular_2012}. These two effects are worth considering as we discuss the possible electrification mechanisms present in our experiments. 

\subsection{Size-dependent bipolar charging}

\begin{figure}
	\centering
	\includegraphics[width=0.6\textwidth]{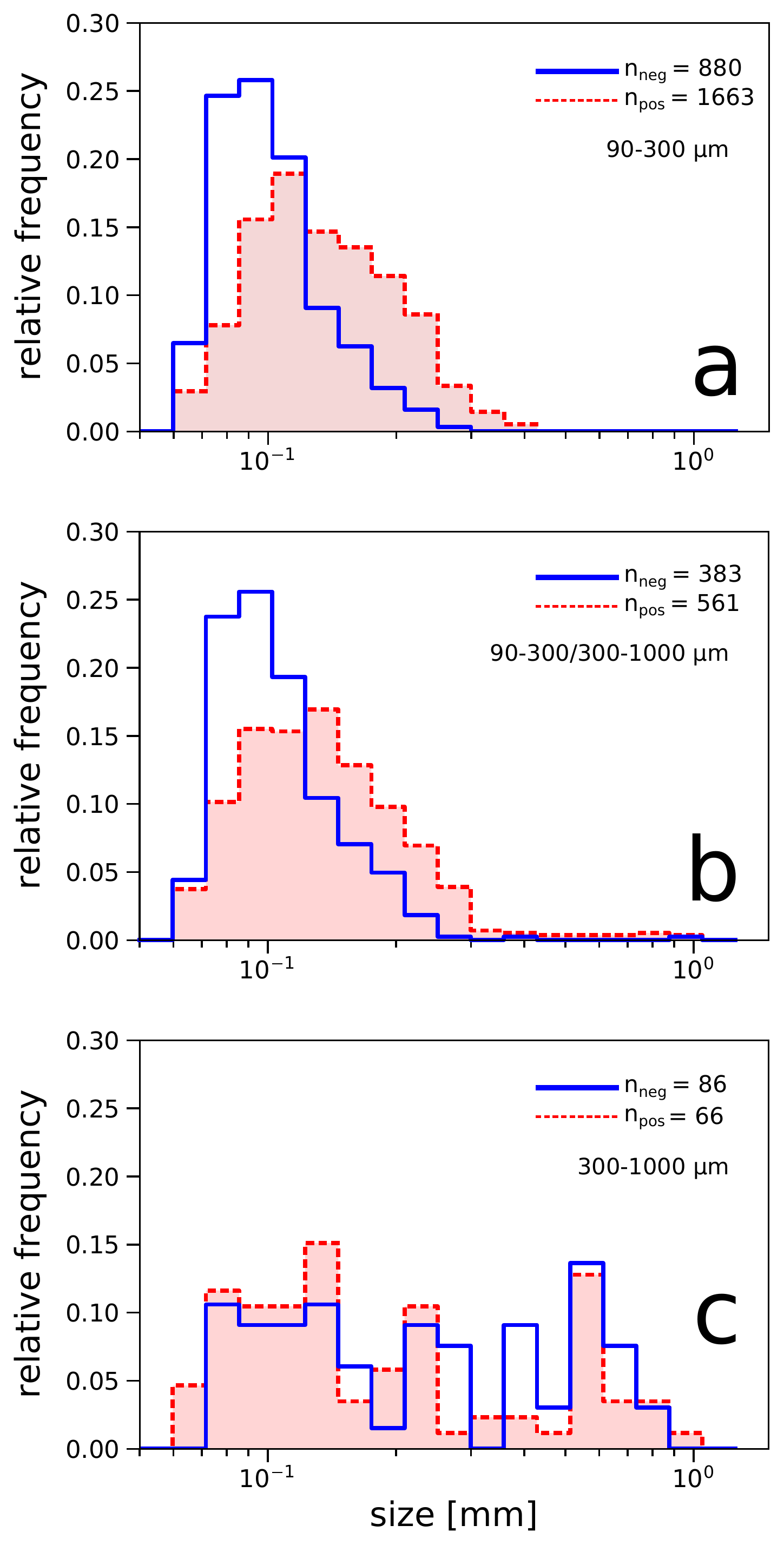}
	\caption{Particle size distributions for positive (red, filled) and negative (blue, empty) particles collected by the  electrostatic separator; a) 90-300 $\mu$m; b) 90-300 and 300-1000 $\mu$m (at 1:1 weight ratio); and c) 300-1000 $\mu$m. Number of positive (n\textsubscript{pos}) and negative  (n\textsubscript{neg}) particles analyzed are reported in the red and blue text respectively. Charge partitioning among particles of different sizes occurs for all granulometries, but is more pronounced for the samples containing small particles (90-300 $\mu$m).}
	\label{histograms}
\end{figure}

Grains leaving the spark-bearing jet fell through the ESS (\textbf{Figure \ref{setup1}b}), where they passed through a strong electric field and were separated by charge polarity. The size distributions for the negative and positive samples for each granulometry as measured using an optical microscope are shown in \textbf{Figure \ref{histograms}} (negative: unfilled, blue curves; positive: filled, red dotted curves). Again, each histogram aggregates data from four separate experiments. The sums of these histograms result in the total post-experiment histograms in \textbf{Figure \ref{histogramsO}}. For both the small and medium distributions, we observe clear size-dependent bipolar charging. Negatively charged particles (blue, unfilled histograms in \textbf{Figure \ref{histograms}}) were more likely to have smaller diameters than positively charged grains (red, filled histograms).  We did not detect statistically significant SDBC in the large sample. We suspect that this absence stems from the sampling limitations discussed above rather than the charging characteristics of the granular sample itself.

This partitioning of charge polarity among particles of different sizes in the small and medium samples is consistent with previous descriptions of triboelectricity in granular materials \citep{zhao2003bipolar,forward_particle-size_2009, waitukaitis2014size, toth2017particle}. As noted by \cite{forward_particle-size_2009}, such size effect appears to be a universal characteristic of chemically-homogeneous triboelectrification. Yet, despite the large number of works that report SDBC, the microphysical processes that cause particles of different sizes to concentrate charge of opposite polarities remain obscure. Some authors have invoked an exchange of trapped electrons between surfaces to account for SDBC \citep{lacks_effect_2007, forward_particle-size_2009}. The trapped electron model proposes that all material surfaces have electrons caught in unfavorable high-energy states. Surfaces are also assumed to have an even larger number of empty, low-energy states. For insulating particles, high- and low-energy states on the same particle surface cannot readily recombine because of the material's low conductivity. When two particles are brought into contact, however, there is a finite probability that a trapped electron on one surface will transfer to a low-energy state on another. Although particles of different sizes have the same density of trapped electrons, the net number of trapped electrons should scale with particle size. Thus, after a comparatively low number of contacts, smaller particles may become depleted in high-energy, donor electrons. Yet, small particles may continue to fill vacant, low-energy states with electrons transferred from large particles.  Over time, there is a net transfer of negative charge from electron-rich, large particles to smaller ones \citep{lacks_effect_2007, forward_particle-size_2009}. 

Despite the trapped electron model's ability to account for the general features of same-material tribocharging, \cite{waitukaitis2014size} have argued that the density of trapped, donor electrons on surfaces may be insufficient to produce the degrees of charging observed in granular media. Recent work posits that, rather than electrons, the main charge carriers exchanged during particle-particle collisions are water ions \citep{gu_role_2013}. This water ion partitioning model brings attention to water films that naturally exist on most surfaces in atmosphere. Within these films, water molecules undergo self-ionization to produce OH\textsuperscript{-} and H\textsuperscript{+}. During grain-grain collisions, some mechanical energy is converted to heat. The temperature of smaller particles climbs faster than that of larger grains. Because H\textsuperscript{+} has a higher mobility than the heavier OH\textsuperscript{-}, positive charge is able to efficiently migrate from a warm, small grain to a large, cool grain. Thus, the water ion partitioning model leads to the same qualitative conclusion as the trapped electron model: small, negative particles and large, positive particles. Ultimately, the mechanisms that bring about SDBC in granular media are an area of active research and it is difficult to point to a single mechanism with any certainty at this time.

While SDBC may be a diagnostic property of triboelectrification, we consider whether such phenomenon occurs in other electrification mechanisms as well. Such attention is warranted because the ESS collected particles smaller than the smallest grains in the original sample (this effect is particularly evident in the experiments with the large granulometry; see \textbf{Figure \ref{histograms}c}). The presence of these fine particles --presumably produced by the disruption of larger particles during the decompression events-- suggests that some amount of fragmentation charging was active during each experiment. In the context of fractocharging, charge partitioning based on size has been reported at least once. By breaking a variety of pumice samples, \cite{james2000volcanic} found that larger and smaller fragments gained opposite charge polarities.  This \textit{fracto-SDBC}, however, varied significantly from that generally observed in frictional electrification. Foremost, \cite{james2000volcanic} found that (in general) larger fragments carried negative charge, whereas small particles charged positively--opposite to that associated with triboelectricity. Additionally, the authors deduced that fracto-SDBC reflects longer, secondary processes --namely, the asymmetric capture of positive ions by particles settling at different velocities-- rather than charge rearrangements occurring during the fracture process itself.  \cite{james2000volcanic} themselves noted that charge polarity segregation is not likely to occur during fragmentation because ``any section of fracture surface has no knowledge of the size of particle to which it is attached." Thus, while fractocharging may efficiently produce elevated charge densities on newly-formed surfaces, the polarity of any given fragment immediately after fracture may involve greater stochasticity. Such behavior stands in contrast to tribo-SDBC, where charge partitioning is believed to occur at the moment two surfaces touch and separate. The expedient nature of SDBC has been demonstrated explicitly in free-fall experiments using high-speed videography \citep{ waitukaitis2014size}. Those investigations reveal charge and charge separation processes occurring on millisecond timescales and imply faster rates for granular media with higher granular temperatures. Conversely, charge partitioning in the fracture mechanism seems to depend on how fast fragments can scavenge ions.  \cite{james2000volcanic} report minute-long timescales for this process. Ultimately, whether or not SDBC is an inherent property of fractoelectric charging (or any other mechanism), as it appears to be in triboelectric ones, remains an open question and should be the focus of future dedicated studies.

\subsection{Charge density on fallout}

\begin{figure}
	\centering
	\includegraphics[width=0.9\textwidth]{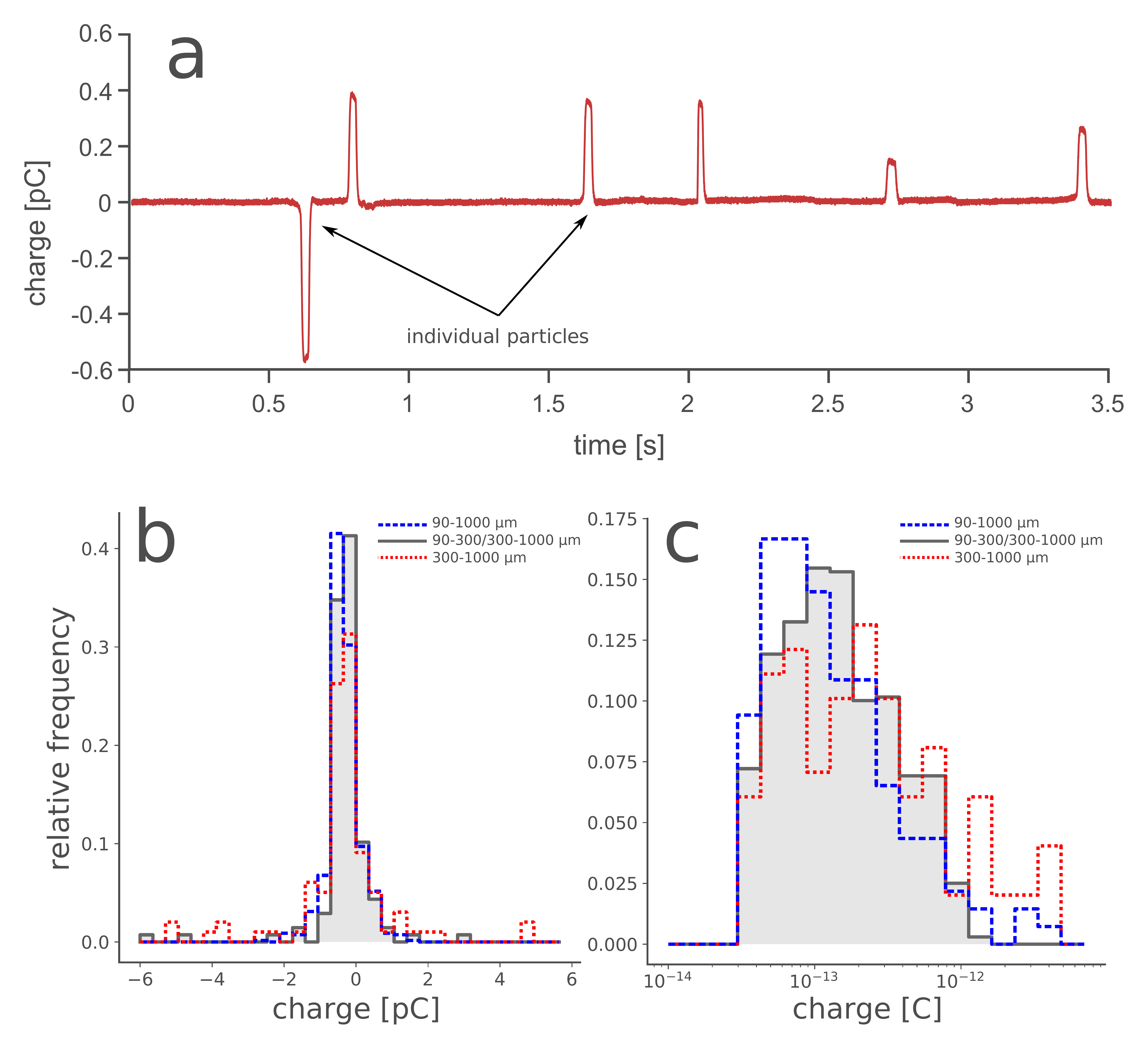}
	\caption{a) Output from one TTFC amplifying stage. Each pulse is representative of the charge on an individual grain traversing the sensing volume. b) The distribution of surface charge on individual particles for each granulometry (dashed blue: 90-300 $\mu$m; solid grey: 90-300/300-1000 $\mu$m; dotted red: 300-1000 $\mu$m). The distributions represent the aggregation of data collected across four experiments. Particles charge both positively and negatively and had mean charges near 0 Coulombs. c) Same as b) but plotted on a logarithmic scale (absolute charge).} 
	\label{charge1}
\end{figure}

The ESS allowed us to determine the distribution of charge polarity on grains falling out of the diverted jet, but not charge magnitude. We used an array of 8 through-type Faraday cups to assess the charge magnitude on particles (\textbf{Figures \ref{setup1}a} and \textbf{\ref{setup2}b}). Particles traversing the sensing volume of a cup produced voltage pulses at the output of an amplification circuit (exemplified in \textbf{Figure \ref{charge1}a}). The amplitudes of the pulses were proportional to the charge on the sampled particles. The histograms in \textbf{Figure \ref{charge1}b} show the distributions of charge on particles in four experiments for each of the three samples. \textbf{Figure \ref{charge1}c} displays the absolute magnitude of that same data on a logarithmic scale. The hard cut-off near 3$\times$10\textsuperscript{-14} Cm\textsuperscript{-2} represents the noise floor of the TTFC. In agreement with the ESS measurements, we find that particles traversing the TTFC array carried both negative and positive charge with a mean charge near zero Coulombs. 

\begin{figure}
	\centering
	\includegraphics[width=0.55\textwidth]{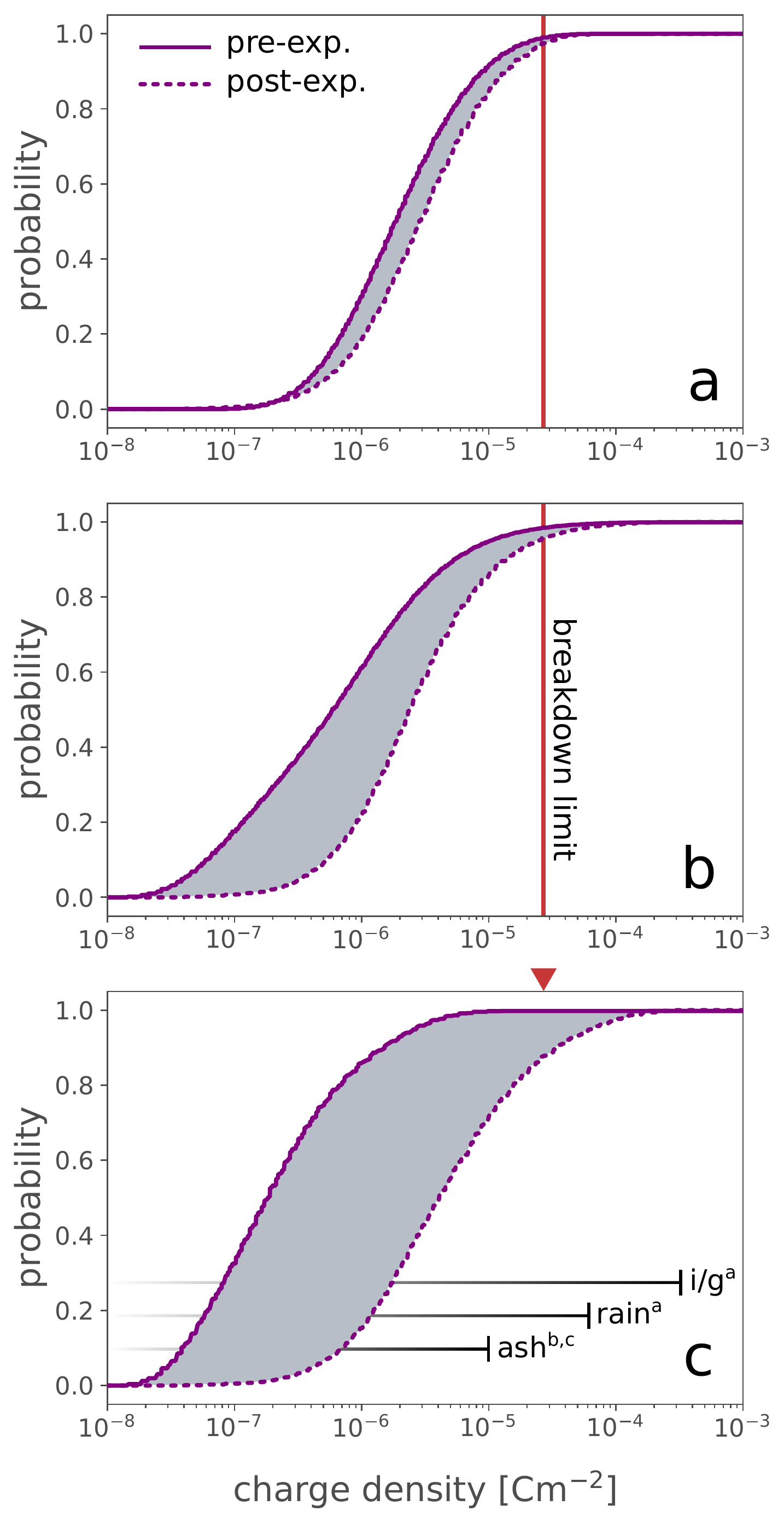}
	\caption{Cumulative probability distributions of surface charge density for the three samples as inferred from the size distributions in \textbf{Figure \ref{histogramsO}} and charge distributions in \textbf{Figure \ref{charge1}}. Note that this surface charge density calculation was done for both the pre- and post-experiment size distributions. a) 90-300 $\mu$m; b) 90-300 and 300-1000 $\mu$m (at 1:1 weight ratio); and c) 300-1000 $\mu$m. The vertical red lines in panels a) and b) and the triangle in panel c) indicate the theoretical maximum surface charge density for large
	surfaces (\textbf{Equation \ref{eq1}}). The horizontal bars in panel c) indicate the ranges of surface charge densities measured on particles in the field: ice and graupel (abbreviated i/g)  \citep{takahashi2012precipitation}, rain \citep{takahashi2012precipitation}, and volcanic ash \citep{gilbert_charge_1991, miura2002measurements}.}
	\label{chargeDen}
\end{figure}

A common way to characterize charging in a granular material is to compute particle surface charge density --that is, the charge normalized by the particle surface area. Because our array of TTFC did not allow for simultaneous measurements of charge and size, we present the surface charge density $\sigma$ as the joint probability given a grain size distribution (using spherical equivalent diameters) and the distribution of measured absolute charges (\textbf{Figure \ref{charge1}c}). The cumulative probability distributions of surface charge density for the three granulometries are shown in \textbf{Figure \ref{chargeDen}}. We compute charge densities using both the pre- and post-experiment grain size distributions (\textbf{Figure \ref{histogramsO}}). In \textbf{Figure \ref{chargeDen}}, the dashed purple curves represent the charge densities computed with the post-experiment granulometery, whereas the solid purple curves are the charge densities computed with the pre-experiment grain size distributions. The grey area between the curves represents the uncertainty involved in these computations. For comparison, \textbf{Figure \ref{chargeDen}c} displays the maximum charge densities measured on ice, graupel (i/g), and rain within East Asian rain bands \citep{takahashi2012precipitation}, as well as ash falling out of a Sakurajima plumes \citep{gilbert_charge_1991, miura2002measurements}. We discuss these comparisons further on.

By invoking Gauss' Law, the theoretical maximum surface charge density on a large surface can be computed as:

\begin{equation} \label{eq1}
    \sigma_{\mbox{max}} = \epsilon_r \epsilon_o E_{\mbox{max}}.
\end{equation}

Above, $\epsilon_o = 8.854 \times$10\textsuperscript{-12} F m\textsuperscript{-1} is the permittivity of vacuum, $\epsilon_r$ is the relative permittivity of gas phase, and E\textsubscript{max} is the breakdown electric field of the gas for a given pressure. We set the relative permittivity equal to 1 and E\textsubscript{max} $= 3\times10^{6}$ Vm\textsuperscript{-1} (corresponding to conditions within the lab). The theoretical maximum charge density  $\sigma_{\mbox{max}}$ is then $2.66  \times$10\textsuperscript{-5} Cm\textsuperscript{-2}. This threshold is indicated in \textbf{Figure \ref{chargeDen}} by the vertical, red lines (panel a and b) and a triangle (panel c). 

Our data shows that the surface charge density on grains spans several orders of magnitude up to the theoretical breakdown limit. However, we find that for all experiments the probability that a particle will have a charge density exceeding this threshold ranges between 2$\%$ and 13$\%$ depending on the grain size distribution (pre- or post-experiment) used to compute the surface charge density distributions. In other words, particles falling out of the diverted plume are overwhelmingly under-saturated in charge according to \textbf{Equation \ref{eq1}}. This apparent under-saturation may result from the fact that we characterized the charge on grains after they transited the region of the jet where discharges occur. As such, we may be measuring only a fraction of the initial charge generated during the decompression event. Indeed, as described in \cite{Mendez2018Infer}, the compressible gas dynamics within the barrel shock may lead to a region of weakened dielectric strength in which particles are forced to shed charge. However, as noted in \cite{gilbert_charge_1991}, particles in dry granular media are rarely observed to carry surface charge densities exceeding 10\textsuperscript{-5} Cm\textsuperscript{-2}. Apparent low surface charge densities on individual grains may simply result from how charge carriers arrange themselves on dielectric surfaces. \cite{baytekin_mosaic_2011} demonstrated that charge may not be distributed uniformly on insulating surfaces. Owing to their high surface resistivities, charged dielectrics can display coexisting regions of high and low surface charge density. Furthermore, these areas need not be of the same polarity. Because a TTFC measures the \textit{net} charge on a particle passing through its sensing volume, the sensor cannot resolve distinct regions of negative and positive charge on grains even if those individual areas have charge densities close to the breakdown limit.

Interestingly (and for comparison), measurements of charge on hydrometeors in thunderclouds reveal that ice, graupel, and rain may carry net charge densities substantially higher than those carried by silicate grains (\textbf{Figure \ref{chargeDen}c}). \cite{takahashi2012precipitation}, for instance, revealed that 500 $\mu$m ice and graupel particles in East Asian rainbands had charges as high as 100 pC, implying surface charge densities in excess of 10\textsuperscript{-4} Cm\textsuperscript{-2}. These elevated particle charge densities indicate that ice-based charging mechanisms in thunderstorms may be more efficient than dusty triboelectrification. 

\section{Implications for electrical processes in volcanic jets}

\label{discussion}

\subsection{Charging mechanisms in the near vent region}

Although size-dependent bipolar charging has been described before in mobilized volcanic ash \citep{hatakeyama_disturbance_1951, kikuchi_atmospheric_1982, miura2002measurements}, our experiments are the first (as far as we are aware) to detect this charge partitioning in spark discharge-bearing granular flows. The presence of SDBC in our shock-tube experiments, designed to replicate the dynamical conditions in the conduit and gas-thrust region, provides direct evidence that tribocharging is a primary electrification mechanism in near-vent volcanic jets and may be responsible for proximal discharges. Although, \cite{behnke2018investigating} found that changes to ambient electric fields are absent during continual RF emissions (indicating that overall negative and positive regions in a jet have not yet formed), \cite{cimarelli_experimental_2014} suggest that centimeter to meter long filamentary discharges would still require some amount of charge separation on small spatial scales. Such charge separation may be driven by turbulent eddies in the flow. Furthermore, this clustering needs to occur on relatively short timescales (a few milliseconds in the case of our shock-tube). Size-dependent triboelectrification meets these criteria given that collisions simultaneously generate charge and distribute charges of opposite polarities among grains of different size \citep{waitukaitis2014size}. Simulations of triboelectricity in granular materials (e.g. \cite{duff_particle_2008}) imply that the speed at which these co-joined processes occur scales with collision frequency. Because of the high particle loading and turbulent kinematics in the gas-thrust region, we suspect that tribocharging and charge separation rates may be very efficient in incipient jets. 

Other proposed ``ice-free" mechanisms may also generate high levels of charge on short timescales within the gas-thrust region. The fragmentation of the magma column or disruptive clast-clast collisions, for example, likely electrify fragments across periods of micro- or even nanoseconds (the duration of active crack propagation). Yet, available experimental data indicates that the segregation of negative from positive fragments happens independently of the fragmentation process and may require substantially longer periods to manifest \citep{james2000volcanic}. Thus, the rapid clustering of negative and positive particles based on size described above may not occur. However, CRF could also be produced by smaller discharges than the proposed 1-10 m sparks in volcanic columns or even the cm-scale discharges in our shock-tube jets. Indeed, \cite{mendez2021detection} have electronically observed spark discharges in granular basalt flows with poor sorting. Unlike the spark discharges in our shock-tube which can be observed directly with an antenna connected to an oscilloscope, the discharges reported in \cite{mendez2021detection} require an amplification stage to be detected. \cite{wurm2019challenge}, working in the context of discharges in Martian dust storms, hypothesize the existence of micro-scale discharges between individual grains or glow discharge between a grain's surface and the atmosphere. These grain-grain discharges would require very little clustering and could be driven directly by the high levels of charging associated with fractoelectrification. Whether or not the collective action of such minute discharges could produce or enhance CRF would be an interesting topic for future research. 

Because our shock-tube does not simulate dynamics beyond the gas-thrust region and each experiment lasts a few milliseconds, we cannot directly comment on any longer charge segregation processes. Nonetheless, field data suggests that these may become important as jets evolve into buoyant plumes. \cite{miura2002measurements} characterized changes in the local electric field during small explosions at Sakurajima. Based on these measurements and assessment of charged fallout, they conclude that Sakurajima plumes consisted of a principal negative charge layer sandwiched between two positive layers. \cite{miura2002measurements} suggest that this positive-negative-positive arrangement comprises a lower region of coarse, positive particles, a middle section of fine, negatively-charged ash, and a top layer of positive gas or aerosols. The charge separation in the two lowermost layers is consistent with  standard SDBC associated with triboelectrification. However, current triboelectric models and our data cannot account for the topmost layer. Such layer may originate from material fracture, as described by \cite{james2000volcanic}, or the decay of radon gas \citep{nicoll2019first}. The low elevations of Sakurajima plumes ($<$ 4 km), as also observed by more recent studies \citep{cimarelli_multiparametric_2016, vossen2021long}, make it unlikely that ice-based charging mechanisms were responsible for observations made by \cite{miura2002measurements}. Together, our experimental data, that of \cite{james2000volcanic}, and field observations suggest that proximal discharges are driven by triboelectrification in conjunction with another mechanism--likely fractoelectrification.  

\subsection{Electrical effects beyond the gas thrust region}

Although our shock-tube experiment does not replicate the physics underlying the generation of near-vent lightning, the measurements of charged particles falling out of a simulated jet allow us to make basic inferences about the electrostatic conditions downstream of the gas-thrust region. Given that near-vent lightning tends to occur immediately after or even concurrently with CRF discharges \citep{behnke2018investigating}, it may be reasonable to assume that both electrical phenomena are driven by the same electrification mechanisms.

\begin{figure}
	\centering
	\includegraphics[width=1.2\textwidth]{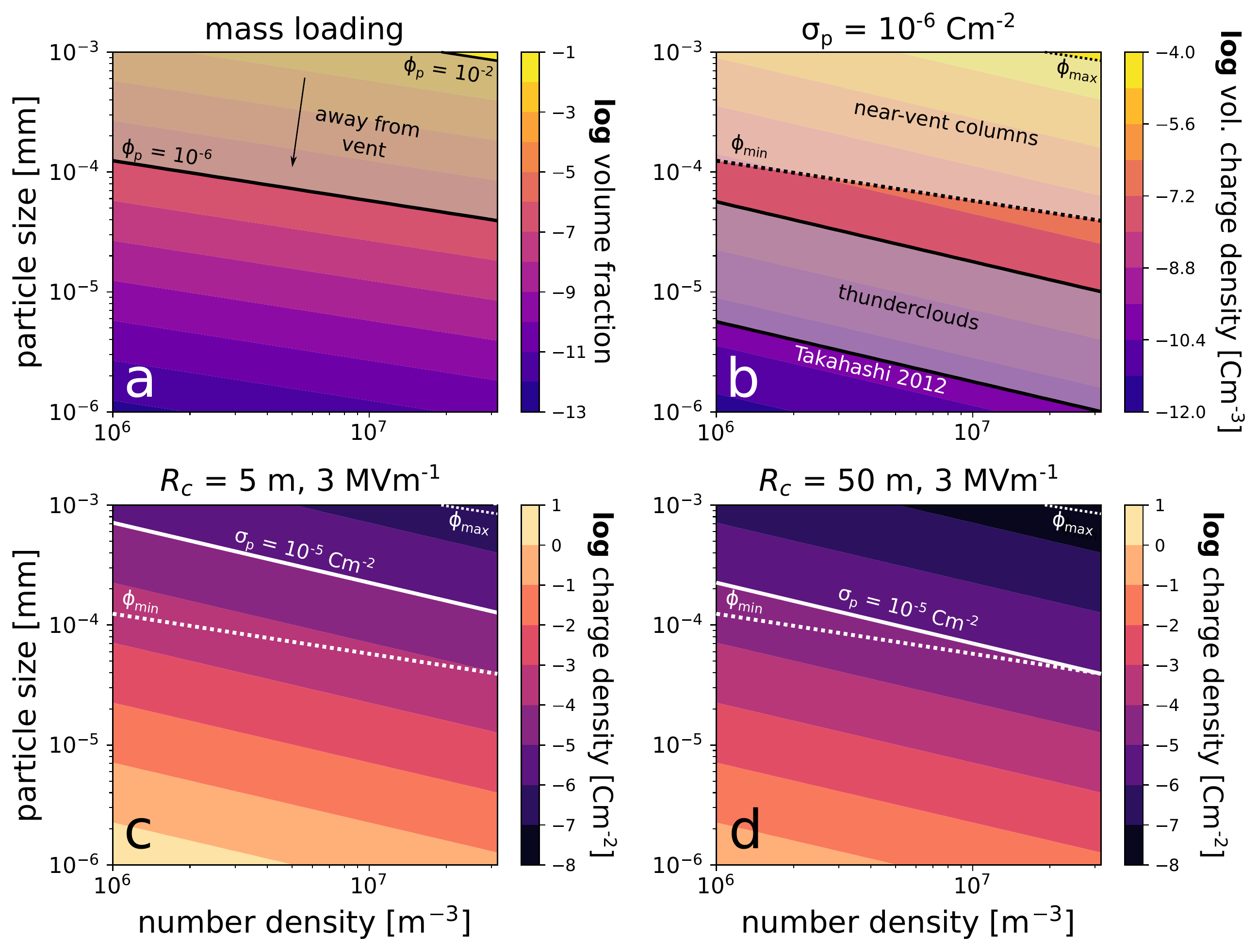}
	\caption{Breakdown criteria for a volcanic ash cloud. a) Mass loading as a function of particle size and particle number density. Shaded region denotes mass loadings expected for proximal columns (see references within text). b) Volumetric charge density as a function of particle size and number density assuming grains with surface charge densities on the order of 10\textsuperscript{-6} Cm\textsuperscript{-2}. Shaded upper region indicates expected volumetric charge densities for proximal columns. Shaded lower area indicates range of volume charge densities measured in thunderclouds \citep{takahashi2012precipitation}. c) Surface charge densities required to produce an electric field at the edge of the region of 3 MVm\textsuperscript{-1} as function of particle size and number density. Region radius set to 5 m. d) Same as c) but for a radius of 50 m.}
	\label{space}
\end{figure}

Observations at Sakurajima reveal that near-vent discharges, with lengths ranging between 10-400 m, are essentially scaled-down equivalents of thunderstorm lightning \citep{aizawa_physical_2016, cimarelli_multiparametric_2016} (we note that, in general, vent-discharges can be longer than those observed at Sakurajima). Unlike CRF sources, near-vent discharges appear to rely on the presence of meso-scale electric fields, indicating the formation of negative and positive regions in a convectively rising column \citep{behnke_changes_2015, behnke2018investigating}.  In their report on Sakurajima near-vent lightning, \cite{aizawa_physical_2016} present a simple model to describe the relationship between flash length and volumetric charge density in these regions. Those authors conclude that near-vent lightning requires small regions of high volumetric charge density. Here, we take this analysis further using the data from our experiments. Assuming each of the charged regions is spherical (radius: $R_r$, radius: $A_r$, volume: $V_r$), we can place first-order constraints on the breakdown conditions given the surface charge densities rendered in \textbf{Figure \ref{chargeDen}}. The electric field $E_s$ at the surface of the region must satisfy Gauss's Law:

\begin{equation} \label{region}
    E_s\oint{dA_r} = \frac{Q(n, A_p, \sigma_p)}{\epsilon_r \epsilon_o}.
\end{equation}

Above, $Q$ is the total charge in the region, which depends on the number density of grains $n$, the total particle surface area $A_p$, and the particle surface charge density $\sigma_p$:

\begin{equation} 
    Q = \sigma_p A_p V_r n.
\end{equation}

Let us consider a region where $n$ is in the range of 10\textsuperscript{6}-10\textsuperscript{8} m\textsuperscript{-3} and the particle diameter is $D_p$ = 10\textsuperscript{-3}-10\textsuperscript{-6} m. This results in volume fractions spanning $\phi$ = 10\textsuperscript{-13} - 10\textsuperscript{-1}, the higher portion of which ($\phi_{min}$ = 10\textsuperscript{-5}, $\phi_{max}$ =  10\textsuperscript{-2}) applies to near-vent volcanic columns (\cite{suzuki2016inter, del2017effect}; see shaded area in \textbf{Figure \ref{space}a}). Assuming any given particle in the region has a charge density on the order of 10\textsuperscript{-6} Cm\textsuperscript{-2} (the typical charge density per particle measured on grains falling out of our experimental jets and that measured on from Sakurajima plumes by \cite{miura2002measurements}), we can compute the region's volumetric charge density as:

\begin{equation} 
    \rho = \pi \sigma_p D_p^2 n
\end{equation}

This calculation is shown in \textbf{Figure \ref{space}b}, where, for comparison, we have also plotted the volumentric charge densities in conventional thunderclouds as measured by radiosondes (\cite{takahashi2012precipitation} and references therein; shaded area). Despite its simplicity, this analytical model is consistent with inferences made by \cite{aizawa_physical_2016} regarding the charge content of Sakurajima columns. Specifically,  we expect the volumetric charge density to be much larger in nascent columns (ranging between 10\textsuperscript{-4} Cm\textsuperscript{-3} near the vent and 10\textsuperscript{-7} Cm\textsuperscript{-3} toward the edge of the gas thrust region) than in conventional thunderstorms (ranging between 10\textsuperscript{-10} and 10\textsuperscript{-8} Cm\textsuperscript{-3}; \cite{takahashi2012precipitation}), despite the fact that individual hydrometeors may carry substantially higher surface charge densities than pyroclasts. 

We can also estimate the surface charge density per particle required to produce the breakdown field $E_S$ at the surface of the region for given region size, particle size, and volume fraction.  Solving \textbf{Equation \ref{region}} for the particle charge density yields:

\begin{equation}
    \sigma_p = \frac{3 E_s \epsilon_r \epsilon_o}{\pi D_p^2 n R_r}.
\end{equation}

For demonstration purposes, let us set the surface breakdown field to the classical ionization limit of $E_s$ = 3 MVm\textsuperscript{-1} (this value would be halved if we are considering two adjacent, oppositely charged regions). We perform these calculations for regions with radii of 5 and 50 m (displayed, respectively, in \textbf{Figure \ref{space}c} and  \textbf{d}), based on the characteristic length of near-vent lightning events  described in \cite{cimarelli_multiparametric_2016} (10s - 100s m). \textbf{Figure \ref{space}c} and \textbf{d} show that for jets with volume fractions in the range of $\phi$ = 10\textsuperscript{-6} - 10\textsuperscript{-2}, the particle charge densities measured in our experiments (10\textsuperscript{-8} - 10\textsuperscript{-5} Cm\textsuperscript{-2})  would be sufficient to produce edge electric fields on the order of MVm\textsuperscript{-1}. We note that maximum electric fields measured in thunderclouds are generally an order of magnitude smaller than the classical limit, suggesting that lightning initiation is controlled by processes other than conventional dielectric breakdown \citep{petersen2008brief}. Assuming similar initiation processes in jets, individual pyroclasts may not need to carry very high net charge densities to meet the lightning initiation criteria if the flow is dense enough (i.e. high mass loading). Additionally, abundant particles may serve to locally enhance a background field, as has been modelled by a number of authors in the context of thunderclouds, reducing the need for extensive, high-magnitude electric fields \citep{dwyer2014physics}.  

As a column matures into a plume, the fate of charge gained in gas thrust region becomes less certain. As noted above, observations at Augustine and Redoubt revealed periods of electrical inactivity following decreases in rates of vent discharges and near-vent lightning \citep{thomas_electrical_2007, behnke_observations_2013, behnke2018investigating}.  One possibility is that as the column entrains air and expands, the volumetric charge density decreases to the point where breakdown conditions are no longer satisfied. Additionally, eruption columns may have be relatively dry during their initial phases. As they cool, however, condensing water, either magmatic or environmental, may promote charge leakage. Although, the partitioning of ions in water monolayers may be required for size-dependent bipolar charging in homogeneous materials (as we have discussed above), \textit{too} much water may have detrimental effects on triboelectrification. Indeed, \cite{stern2019electrification} have shown that the number of discharges in a simulated, oversaturated volcanic granular flow decrease with water content (up to 27 wt$\%$). More recently, \cite{mendez2020microphysical} showed that even smaller amounts of water (below the saturation limit) can reduce the magnitude of triboelectric charging if pyroclasts have high residence times in humid environments. Using a fludizied bed, \cite{toth2017particle} showed that granular materials still charge at high humidities, but that these conditions effectively nullify SDBC (i.e. a large particle is equally likely to be charged negatively as a small particle). Taking these results into consideration, our work suggests that ice-free charging may be extremely efficient at producing charged grains and separating positive and negative particles from each other in relatively dry flows (that is, during the initial, hot phase of an eruption or non-hydromagmatic eruptions). However, the condensation of water may rapidly shut off triboelectric or fractoelectric processes in volcanic columns by reducing both the magnitude of charging and the degree to which charge polarity is separated by size. In this respect, further experimental constraints are needed to better determine the role of water (and other volatiles) on the size-dependent bipolar charging of volcanic materials. 

\section{Conclusions}

We have simulated volcanic jets and proximal electrical phenomena using a shock-tube. To the best of our knowledge, this work demonstrates the presence of size-dependent bipolar charging in a spark-bearing granular flow for the first time. The segregation of charge based on particle size has been associated with triboelectric charging in numerous granular flows over the last 30 years. The detection of SDBC in our experiments, together with investigations at Sakurajima, suggests that frictional electrification may play a significant role in driving electrostatic phenomena within the proximal region of a volcanic column. Specifically, triboelectrification, through SDBC, can both efficiently electrify pyroclasts and drive charge separation to set up the electric fields needed for the production of lightning. Our analysis, however, does not rule out other synergistic electrification mechanisms such as fractocharging or radioactive decay. In fact, field measurements suggest that triboelectrification alone cannot account for observed charge structures in small volcanic columns.   Additionally, we showed that particles emerging from the region where filamentary discharges occur still carry some charge. A first order assessment indicates that particles exiting the gas-thrust region with only a fraction of the theoretical maximum surface charge density would still produce regions with volumetric charge densities capable of sustaining meso-scale lightning. Thus, both vent discharges and larger near-vent lightning may be underpinned by the same electrification mechanisms. 

\section{Acknowledgments}

Joshua M\'endez Harper: Conceptualization, Conducted experiments, Hardware, Data Analysis, Writing. Corrado Cimarelli: Conceptualization, Conducted experiments, Reviewing and Editing. Valeria Cigala: Conducted experiments. Ulrich Kueppers: Conducted experiments. Josef Dufek: Conceptualization, Writing- Reviewing, and Editing.

C.C. acknowledges the
support of Deutsche Forschungsgemeinschaft project CI 254/2-1. J.M.H and J.D. acknowledge the support of NASA SSW 80NSSC19K1211 and NASA IDS 1911057Z4.


\end{document}